\documentclass[a4paper,11pt]{article}
\pdfoutput=1 

\usepackage{jinstpub} 
\usepackage{ucs}
\usepackage[utf8x]{inputenc}
\usepackage[T1]{fontenc}
\usepackage{siunitx}
\usepackage{multirow}
\usepackage{ragged2e}
\usepackage{tabularx}
\usepackage{booktabs}

\newcolumntype{L}{>{\raggedright\arraybackslash}X}
\newcolumntype{C}{>{\centering\arraybackslash}X}

\title{\boldmath Architecture of the data aggregation and
streaming system for the European Spallation Source neutron instrument suite}

\author[a,1]{A. H. C. Mukai,\note{Corresponding author.}}
\author[a]{M. J. Clarke,}
\author[a]{M. J. Christensen,}
\author[a]{J. M. C. Nilsson,}
\author[a]{M. G. Shetty,}
\author[b]{M. Brambilla,}
\author[b]{D. Werder,}
\author[b]{M. Könnecke,}
\author[c]{J. Harper,}
\author[d]{M. D. Jones,}
\author[c]{F. A. Akeroyd,}
\author[e]{C. Reis,}
\author[e]{G. Kourousias}
\author[a]{and T. S. Richter}

\affiliation[a]{European Spallation Source ERIC,\\Ole Maaløes Vej 3, 2200
Copenhagen N, Denmark}
\affiliation[b]{Paul Scherrer Institut,\\5232 Villigen PSI, Switzerland}
\affiliation[c]{ISIS Neutron and Muon Source, Science and Technology Facilities
Council,\\Rutherford Appleton Laboratory, Didcot, OX11 0QX, United Kingdom}
\affiliation[d]{Tessella,\\26 The Quadrant, Abingdon Science Park, Abingdon,
Oxfordshire, OX14 3YS, United Kingdom}
\affiliation[e]{Elettra Sincrotrone Trieste,\\Strada Statale 14, km 163.5, 34149
Basovizza, Trieste, Italy}

\emailAdd{afonso.mukai@esss.se}

\abstract{
  The European Spallation Source (ESS) will provide long neutron pulses for
  experiments on a suite of different instruments. Most of these will perform
  neutron data acquisition in event mode, i.e.\ each detected neutron will be
  characterised by one absolute timestamp and pixel identifier pair. Slow
  controls metadata from EPICS, such as sample environment measurements and
  motor positions, will also be timestamped at their source, so that all data
  and metadata are streamed as a list of events instead of histograms. A
  flexible data aggregation and streaming system is being developed combining
  both open source third-party software and in-house development. This is to be
  used at ESS and other neutron scattering facilities like ISIS and SINQ,
  replacing legacy solutions by a shared software collection maintained by a
  cross-facility effort. The architecture of the Apache Kafka-based system, its
  metadata forwarding and NeXus file writing components are presented, along
  with test results demonstrating their integration and the scalability in terms
  of performance.
}


\begin{document}
\maketitle
\flushbottom

\section{Introduction}


The European Spallation Source (ESS) is a spallation neutron source under
construction in Lund, Sweden. A linear accelerator produces a high intensity
proton beam which is guided on a tungsten target, with the spallation process
generating fast neutrons from the heavy tungsten atoms. These fast neutrons are
slowed down in a moderator and then guided onto samples with which they will
interact. From the interaction of samples with neutrons, scientists can learn
about the properties of materials. Applications include fundamental solid state
physics, materials science, crystallography, biology and archaeology.

ESS will operate as a user facility: external users will come to perform neutron
scattering experiments using the provided high brightness, long pulse neutron
beam. A suite of instruments covering different techniques, such as imaging,
small angle scattering, reflectometry, diffraction and spectroscopy, will be
able to use the unprecedented flux there for experiments \cite{ess:tdr}.

Due to the high brilliance neutron beam, high performance computing is needed
for data processing. Located in Copenhagen, Denmark, the ESS Data Management and
Software Centre (DMSC) develops, maintains and operates a software and hardware
infrastructure for tasks ranging from data acquisition, aggregation and
streaming to reduction and analysis, both in real time and offline. The DMSC
works in close collaboration with in-kind and project partners across Europe,
including Paul Scherrer Institut (PSI) in Switzerland, the Science and
Technology Facilities Council (STFC) in the United Kingdom, and Elettra in
Italy.


Most ESS instruments will acquire neutron data from detectors in event mode,
i.e.\ as a list of pixel number-timestamp pairs. Slow metadata obtained from
devices using the Experimental Physics and Industrial Control System (EPICS)
also come in the form of timestamped values \cite{website:epics}. An aggregation
and streaming software architecture based on Apache Kafka \cite{website:kafka}
will make these data available to applications performing tasks such as file
writing using the NeXus format \cite{nexus:article} and live data reduction with
feedback to the experiment control as well as visualisation.


Due to the high brightness of the ESS neutron beam, the solutions for data
acquisition must be able to cope with correspondingly high data rates, making
performance an important requirement, as will be discussed in this article.
Considering the future possibilities of expansion of the ESS instrument suite
and of upgrades to initial instruments, good scalability of the solution is
essential. The selected open source components and the design of the in-house
developed software take this into consideration.


In this article, the overall architecture of the data aggregation and streaming
system is described, with a focus on the core components: Apache Kafka, the
EPICS to Kafka Forwarder and the NeXus File Writer. Section~\ref{sec:event}
discusses data acquisition in event mode, the requirements arising from it and
also the ESS approach to timing and timestamping. In section~\ref{sec:arch}, the
architecture and components are introduced and discussed. The current deployments
of the system as well as integration and performance scalability tests are
presented in section~\ref{sec:results}, with conclusions summarised in
section~\ref{sec:conclusion}.

\section{Event mode acquisition}
\label{sec:event}

\begin{figure}[htb]
  \centering
  \includegraphics[width=0.9\textwidth]{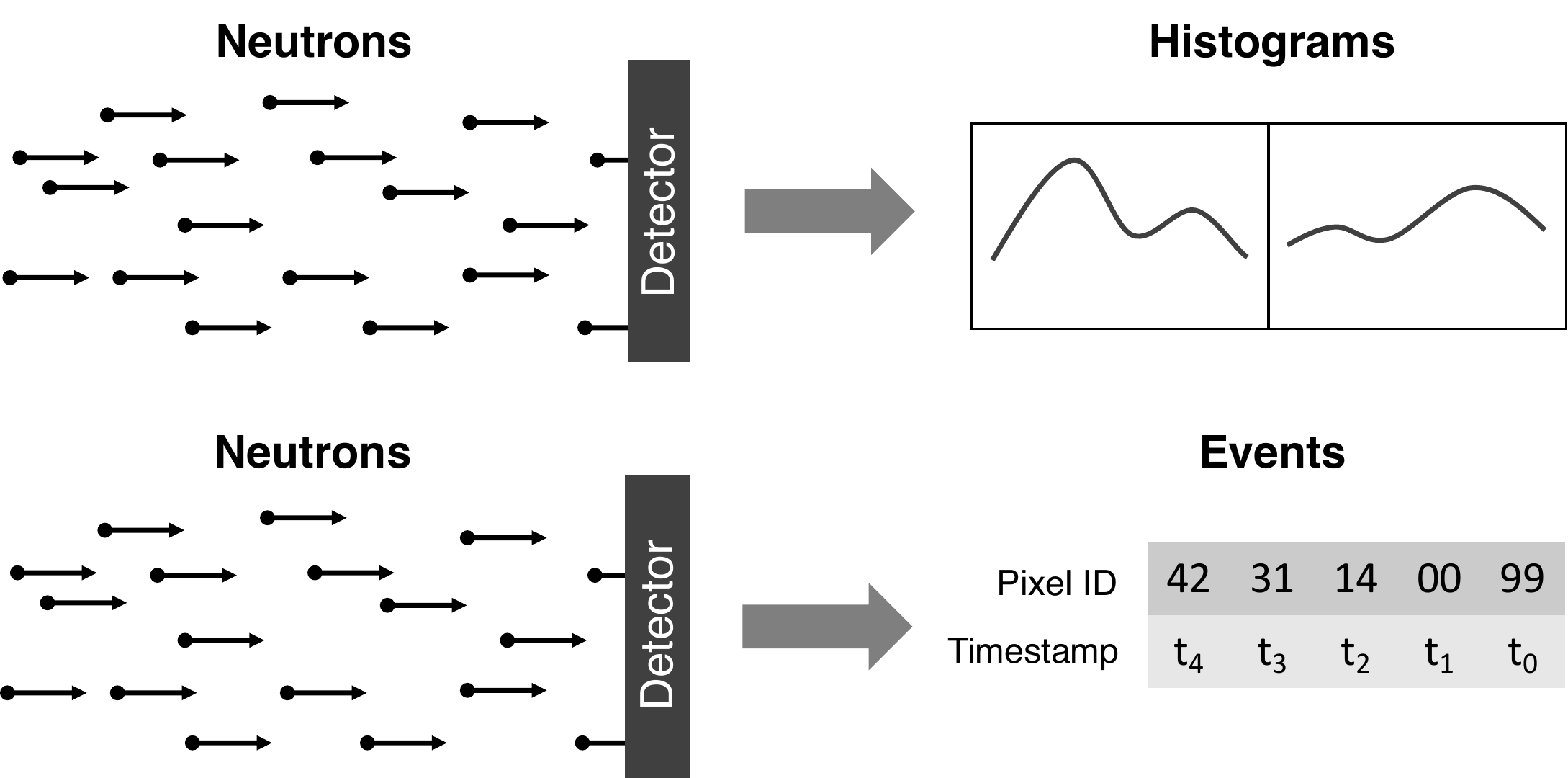}
  \caption{Data acquisition in histograms and event mode. In the first mode,
  counts from neutrons incident on the detector are accumulated and sent as a
  histogram. In the latter, pairs of timestamp and pixel identifier are sent for
  each detected neutron in a list of individual events.}
  \label{fig:histograms_vs_events}
\end{figure}

Acquisition of neutron instrument data is frequently done in histogram mode, in
which detector data over a region is acquired for an interval and presented as a
set of accumulated counts. In event mode acquisition, each neutron count on the
detector generates an individual event consisting of a pixel number, which
uniquely identifies the location where the neutron hit the detector, and a
timestamp. Acquired data are sent forward as a sequence of events.
Figure~\ref{fig:histograms_vs_events} illustrates the difference between these
two modes of data acquisition.

Postponing histogram generation provides additional flexibility to the
data acquisition process, as the conversion of neutron detection events into a
histogram results in loss of information (the individual timestamps are lost and
the final histogram only gives information as detailed as the binning and
integration time allow). Not only can the data in an event list later be binned
in any desired resolution, individual (sets of) events can also be filtered out
depending on metadata parameters like, for example, sample environment
information. In ESS, event mode acquisition will occur with no hardware veto,
that is, data acquisition will not be automatically inhibited should a neutron
chopper be out of phase. As chopper metadata will be recorded, filtering
can be done by discarding events for intervals where choppers were out of phase,
or by taking into account the instantaneous chopper phases, as well as source
and target parameters and correcting the neutron wavelengths for it.

\begin{figure}[htb]
  \centering
  \includegraphics[width=\textwidth]{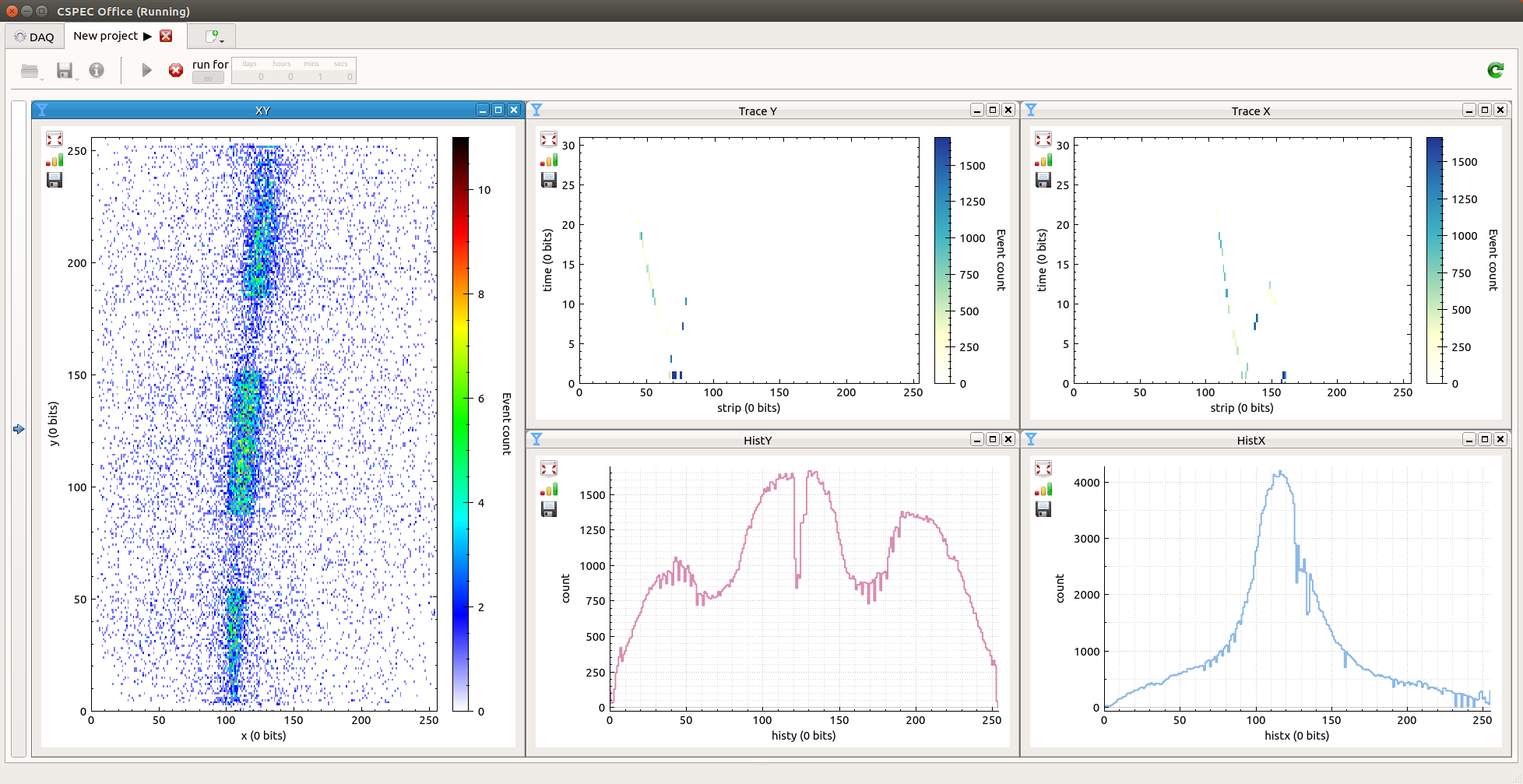}
  \caption{Screenshot of the detector commissioning tool Daquiri with detector prototype test data. The two smaller plots on the top show event traces (for a single event) before processing by the EFU. The plot on the left shows event counts on a detector surface after processing by the EFU.}
  \label{fig:daquiri_screen}
\end{figure}

The data aggregation and streaming system will receive neutron data and metadata
from different sources. Event data come from Event Formation Unit (EFU)
software running on servers that interface with the detector readout electronics
and obtain the
pixel identifier-timestamp pair from the raw detector data~\cite{efu:article}. Figure~\ref{fig:daquiri_screen} shows a screenshot of the detector commissioning tool Daquiri~\cite{sw:daquiri} with examples of event data before and after processing by the EFU. EPICS metadata will be sent by the EPICS to Kafka Forwarder
application, which subscribes to value updates from multiple Input/Output
Controllers (IOCs) that publish the readings for beamline devices like
temperature controllers, motors and choppers. The highest data rates are
expected to arise from neutron event data; these will be discussed next,
followed by the ESS approach to timing and timestamping.

\subsection{ESS instrument data rates}


As a result of the high neutron beam brightness, the expected event data rates
for ESS instruments are correspondingly high. Each detector pixel identifier and
timestamp will be encoded in a pair of 32-bit numbers \cite{sw:schemas}; using
these values, table \ref{tab:detector_rates} presents estimates of the data
rates expected to be sent from the EFUs into the aggregation and streaming
system \cite{efu:article}.

\begin{table}[htb]
  \centering
  \caption{Anticipated neutron and detector event rate estimates for some early ESS
    instruments.}
  \begin{tabularx}{115mm}{LCC}
    \toprule
    \textbf{Instrument} & \textbf{Global Average Rate} & \textbf{Data Rate} \\
                        & [MHz]                        & [MB/s]             \\
    \midrule
    C-SPEC              & 10                           & 80                 \\
    ESTIA               & 500                          & 4000               \\
    FREIA               & 100                          & 800                \\
    LoKI                & 37                           & 300                \\
    NMX                 & 5                            & 40                 \\
    SKADI               & 37                           & 300                \\
    T-REX               & 10                           & 80                 \\
    \bottomrule
  \end{tabularx}
  \label{tab:detector_rates}
\end{table}


As the table shows, expected rates cover a range of almost three orders of magnitude,
from \SI{10}{MB/s} to \SI{4}{GB/s}. In addition to the bare detector neutron rate, the metadata
from beam monitors, choppers, (fast) sample environments and so on also need to be
considered. It is expected that the total metadata volume can be up to the order of the
neutron data for an average instrument. This sets the requirements for the rates the
system must support for online visualisation and file writing. The accelerator
will be ramped up in power over a few years and
instruments will be built and commissioned over a period of time, so ESS will not
initially operate at maximum rates; the aggregation and streaming solution
should therefore be scalable, allowing it to expand by adding more resources as necessary.

\subsection{Timing system and timestamping of data}


For devices that produce data or metadata, all readout values will be timestamped at
the source (with deterministic latency) using information from the ESS timing system.
In this system a tree topology of event receivers with an event generator at the root
is used for distributing trigger and clock signals,
timestamps, and accelerator parameters across
the facility with deterministic and compensated latency. In normal operations, the ESS neutron pulses
of \SI{2.86}{ms} duration will have a repetition rate of \SI{14}{Hz}. The timing
system will make available beam parameters such as the pulse length and proton
beam energy and will be integrated with EPICS \cite{esstiming:conf}.


Neutron detector data acquisition in event mode relies on information that is
provided by the ESS timing system, such as accurate timestamps for neutron
events coming from the detector readout electronics and accelerator pulse
sequence numbers. Events are accumulated and sent in a serialised message (see Section \ref{sec:schemas}) that includes a 64-bit pulse time in nanoseconds since an epoch and an array of 32-bit neutron timestamps in nanoseconds from pulse time. Furthermore, to be able to make sense of data and process the
events, sample conditions must be known at any given instant; for that, the
timestamped metadata are used, using values provided by EPICS process variables.
As all the timestamping happens at the data source, later transport over normal
computer networks, software data aggregation and processing in standard
non-realtime systems and so on do not distort the data by adding latency.
Latency times for the data must be kept reasonably low to
allow live experiment data visualisation by the user and enable useful
automatic feedback from live data processing to the control of the
ongoing experiment.

\section{Data aggregation and streaming}
\label{sec:arch}


Data acquisition in event mode produces a stream of pixel identifier-timestamp
pairs, which are subsequently processed by other applications and combined with
metadata to enable a user to make sense of the experiment being performed.
By selecting a system
architecture based on aggregation and streaming~\cite{streaming:conf}, the data
publishers do not have to be directly connected to every data subscriber, thus
decoupling the two categories and allowing them to scale separately. A common
serialisation format is needed to ensure the different applications all
understand the data in the same way.

\begin{figure}[htb]
  \centering
  \includegraphics[width=0.9\textwidth]{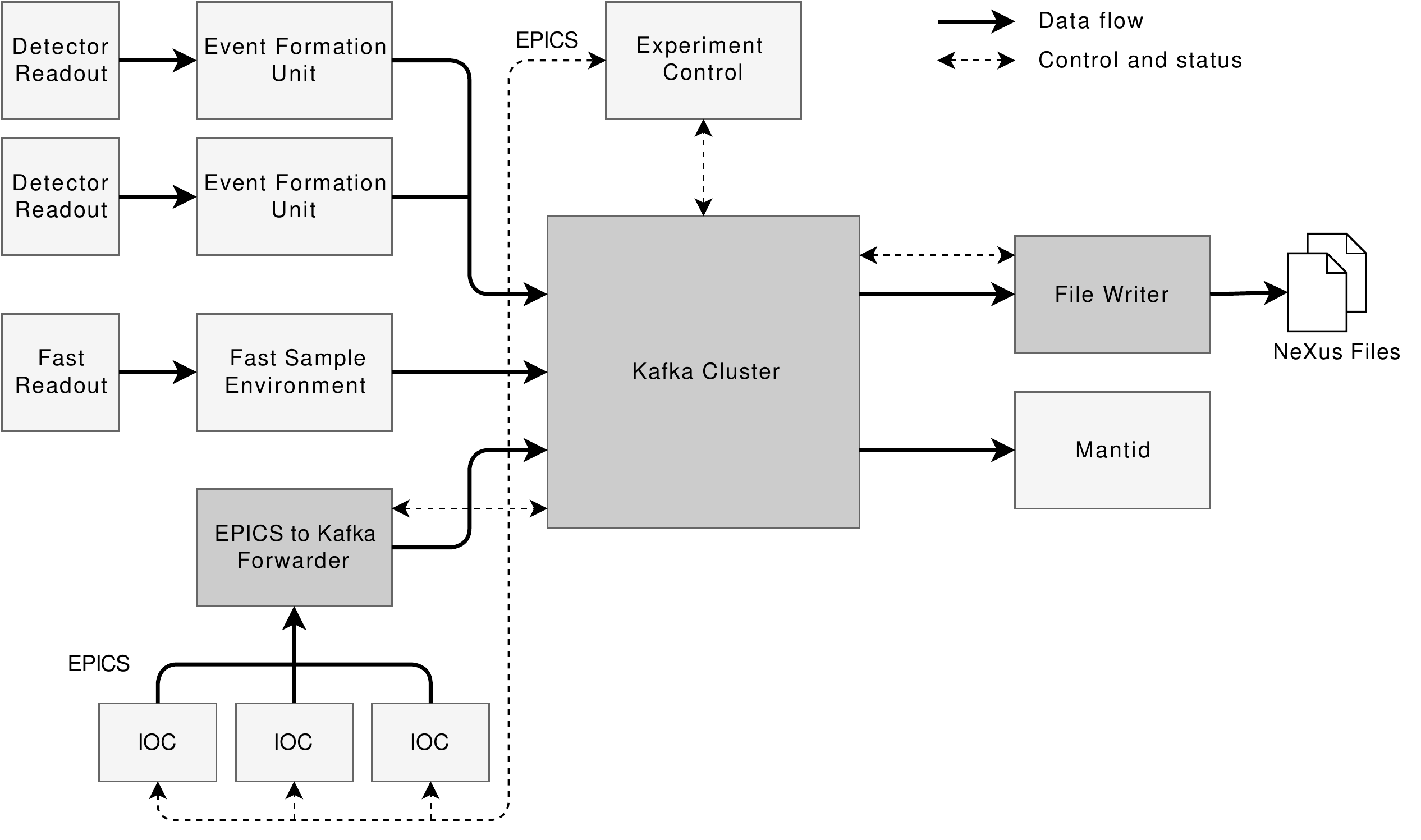}
  \caption{System architecture overview. Apache Kafka provides the central
    infrastructure that connects data producers and consumers. Components in
    dark grey are the main focus of the
    data aggregation and streaming and are described
    in this publication.}
  \label{fig:system_arch_overview}
\end{figure}

An overall view of the adopted system architecture is presented in figure
\ref{fig:system_arch_overview}. The data sources depicted in that diagram
are the Event Formation Units; the Fast Sample
Environment, which sends event data from fast readout systems such as
alternating electric and magnetic fields; and the EPICS to Kafka Forwarder,
which sends experiment metadata from EPICS IOCs. Subscribing to these data are the
File Writer, which generates files conforming to the NeXus format
\cite{nexus:article} for offline reduction and analysis, and the data reduction
application Mantid \cite{mantid:article} performing live data reduction to
enable online visualisation. The streams of data are brokered through an Apache
Kafka cluster, which is also used for sending and receiving control and status
messages. In the next sections the
central components for aggregation and streaming are discussed in more detail.

Instead of using specifically-tailored solutions for the ESS, the data aggregation and streaming system relies on a number of open source tools and libraries, and the applications developed in house are also generic and configurable.
While specific solutions can be optimised to minimise resource usage considering detailed information about the systems they cater to, using more generic tools provides leverage through a larger developer and user community, alleviating the local development and maintenance work that can focus on domain-specific applications.
The decoupling of the system components combined with the common serialisation
format allows different facilities to plug in their own specific applications,
while using the other parts of the infrastructure. An example of such use of the
system by the ISIS Neutron and Muon Source is presented in section
\ref{sec:deployments}.

\subsection{Apache Kafka}
\label{sec:kafka}


Apache Kafka is an open source distributed streaming platform that provides a
scalable solution for applications to publish and subscribe to streams of data
\cite{website:kafka}.
Kafka has been selected as the central technology for aggregation and
streaming based on its rich set of implemented functionality, scalability,
large and active user and developer communities, and the availability of good
documentation~\cite{brightness:d51}.

In Kafka
terminology, the \emph{producers} (publishers) write data to named
channels known as \emph{topics}, to which
\emph{consumers} can subscribe. A cluster is formed by a number of Kafka
brokers, with the possibility of partitioning topics across different brokers
for scalability and replicating them for availability in case of broker
failure. Data are persisted to disk for a configurable time or storage limit in
the cluster, allowing consumers to obtain historical data. The cluster rate
capacity can be increased by adding more brokers and partitioning topics across
them.

\begin{figure}[tb]
  \centering
  \includegraphics{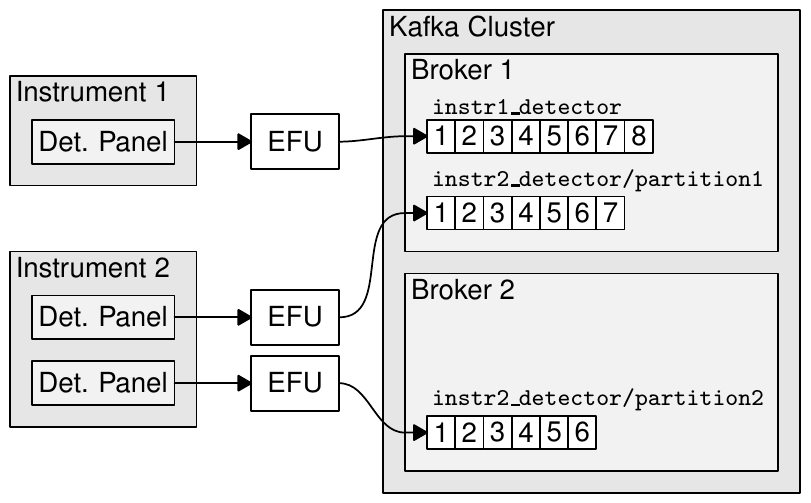}
  \caption{Neutron detector data from Event Formation Units sent to different
    Apache Kafka topics and partitions. Partitioning of topics across different
    Kafka brokers provides scalability to handle high data rates, as each broker
    runs on a server with dedicated network interfaces and storage resources.}
  \label{fig:efu_kafka_topics_and_partitions}
\end{figure}

Figure
\ref{fig:efu_kafka_topics_and_partitions} illustrates how topic partitioning can
be used to allow detector readout to scale by writing parts of the event streams
to different Kafka brokers. It also shows that the intention is to use one common
Kafka installation for all instruments. This shared system can
cope with occasional peak loads from instruments far above their average rate,
without needing to over supply capacity at every instrument. In addition to
the depicted scaling out, Kafka will also be used to provide redundancy by
duplication of the topics for the
raw data storage, so experimental data is safely persisted early on.

\begin{figure}[tb]
  \centering
  \includegraphics{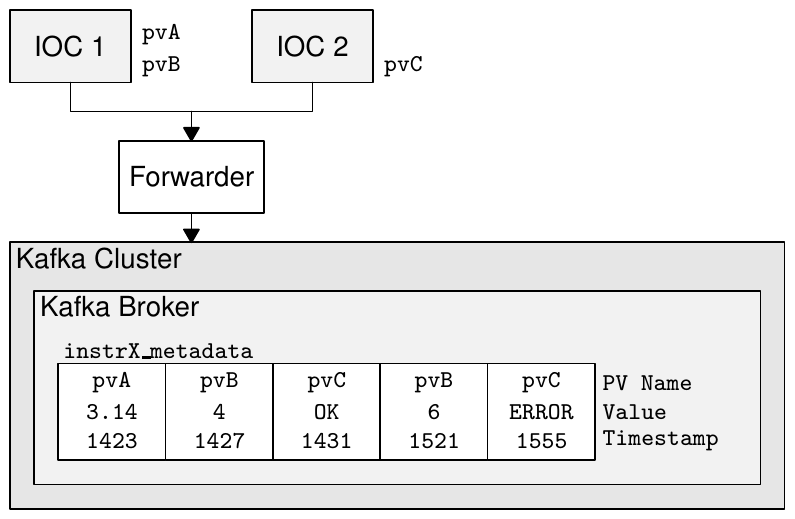}
  \caption{EPICS to Kafka Forwarder multiplexing data from different EPICS
    process variables (PVs) using the same Kafka topic.}
  \label{fig:forwarder_kafka_topic_multiplexing}
\end{figure}

Fast sample environment data will also be sent to
the cluster using an EFU-based solution.
For lower rate data sources, such as EPICS metadata, whose values are updated
at rates in a typical range of \SI{0.1}{Hz} to \SI{100}{Hz}, the same Kafka
topic can be used for multiple sources. The EPICS to Kafka Forwarder serialises
and multiplexes different sources over a single topic, as illustrated in figure
\ref{fig:forwarder_kafka_topic_multiplexing}.

\subsection{Serialisation and schemas}
\label{sec:schemas}


As the aggregation and streaming system is comprised of distributed producers
and consumers that exchange data and commands through Apache Kafka, a
well-defined common set of schemas is required to ensure the applications can
make sense of the data being exchanged. Google FlatBuffers, an efficient
cross-platform serialisation library \cite{website:flatbuffers}, has been chosen
for serialising neutron event data and beamline metadata. A common Streaming
Data Types repository with an agreed procedure for
introducing changes is used for controlling the schemas that
define the message format \cite{sw:schemas}. Schemas can include other schema files, a feature
that is used in the neutron event schema to add facility-specific information to
them.

Serialised messages are not self-describing (i.e.\ they cannot be deserialised
without knowledge of the schema), but include a four-byte file identifier;
this is used to identify the schema required to deserialise the message by
the consuming application. A convention of prefixing the schema file name with
the identifier has been adopted to help identify the files in the repository. The
FlatBuffers library allows schema fields to be optional or required, and fields
can be added and deprecated. FlatBuffers provides a compiler
that takes schemas as input and generates source files in
a number of programming languages, including C++ and Python, the main
languages used by the ESS aggregation and streaming programs.
These source files provide language-specific methods for serialising and deserialising
messages to and from buffers according to a given schema.

The schema used by the EFU for sending event data to the Kafka cluster includes a global timestamp associated with a neutron pulse and a pair of arrays containing event timestamps and pixel numbers; buffer size is dominated by the number of events in it.
For message sizes of about \SI{1}{MB} (see figure \ref{fig:kafka_perf} and the discussion in section \ref{sec:perf}), this means hundreds of thousands of events can be accumulated by the EFU in each message.
On high data rate instruments, buffers can be filled with events and sent after a certain message size has been reached, while for lower rate instruments messages can be sent for each pulse, thus balancing throughput and latency concerns.


Apache Kafka is also used for the exchange of control and status messages
between applications. Examples of control messages include commands for starting
and stopping file writing by the Experiment Control Program, and both the EPICS
to Kafka Forwarder and the NeXus File Writer send status messages reporting
their state and activity periodically. These messages are currently sent in JSON format to
dedicated Kafka topics, with a low rate of the order of \SI{1}{Hz}.

\subsection{EPICS Forwarder}
\label{sec:forwarder}

The EPICS to Kafka Forwarder~\cite{sw:forwarder} is the component in the data
aggregation and streaming architecture that monitors a list of EPICS process
variables and makes these updates available for the rest of the components via
the Kafka cluster.

Monitoring of the process variables is done using EPICS version 4, which
supports both the older Channel Access protocol and the
newer pvAccess \cite{website:epicsv4}. The list of monitored variables can be
set in a configuration file that is read by the EPICS to Kafka Forwarder at
startup and can also be modified via JSON command messages, which are delivered
through Kafka.

One or more conversion modules can be attached to
each monitored process variable, which take the responsibility to convert the
received EPICS value into a buffer for the desired schema, as described in
section~\ref{sec:schemas}. It is also possible
to attach an optional conversion module to several monitored process variables to cater
for more elaborate setups.
The converted update is delivered
to the configured topic in the Kafka cluster (see section~\ref{sec:kafka})
via the C/C++ librdkafka
library~\cite{website:librdkafka}, which handles the network connections,
buffering and delivery reports.

The EPICS to Kafka Forwarder features a highly parallel design based on worker
queues and a configurable number of worker threads. It is designed to handle
updates at a high frequency, which is achieved by keeping the EPICS update
callback as lightweight as possible, while still providing the functionality to
enable concurrency in the later stages of the forwarder pipeline. The update of
the process variable can go through the pipeline stages while the thread that
invokes the monitor callback, which is owned by the EPICS library, has already
returned control to the library.

At runtime, the forwarder sends status and statistics about the
monitored process variables to the Kafka cluster in JSON format. This
information is useful for system monitoring
by the Experiment Control Program and for diagnostics.

\subsection{Writing data to file}

While the Kafka cluster partitions persist the
full stream of experimental data in roughly temporal order as they are
measured, it is beneficial for the performance of offline data reduction or
analysis algorithms to
be able to access the data in larger homogeneous chunks. File-based storage
is also well established in computing architectures and in the scientific community
with agreed, self describing formats that can be a long term scientific record
for experiments at ESS.

NeXus is a common data format for neutron, X-ray and muon science
\cite{nexus:article} and is based on the Hierarchical Data Format 5 (HDF5)
\cite{nexus:hdf5}. Many institutions and existing pieces of software use NeXus. By opting to
store data in this format the ESS benefits from existing support in software,
including Mantid \cite{mantid:article}.
The NeXus File Writer~\cite{sw:filewriter} creates HDF5 files that contain a
configured set of the available data and metadata from experiments.

HDF5 stores data in a format analogous to a filesystem. Directory-like objects
called \emph{groups} contain child groups and \emph{datasets}. Datasets are
analogous to files and contain a scalar value or a multidimensional array of
data which may also be compressed; datasets and groups can each have
\emph{attributes} containing supporting metadata. Links can be used in place of
groups or datasets and can link to locations in the same file or in external
files.

\begin{figure}[htb]
  \centering
  \includegraphics[scale=0.5]{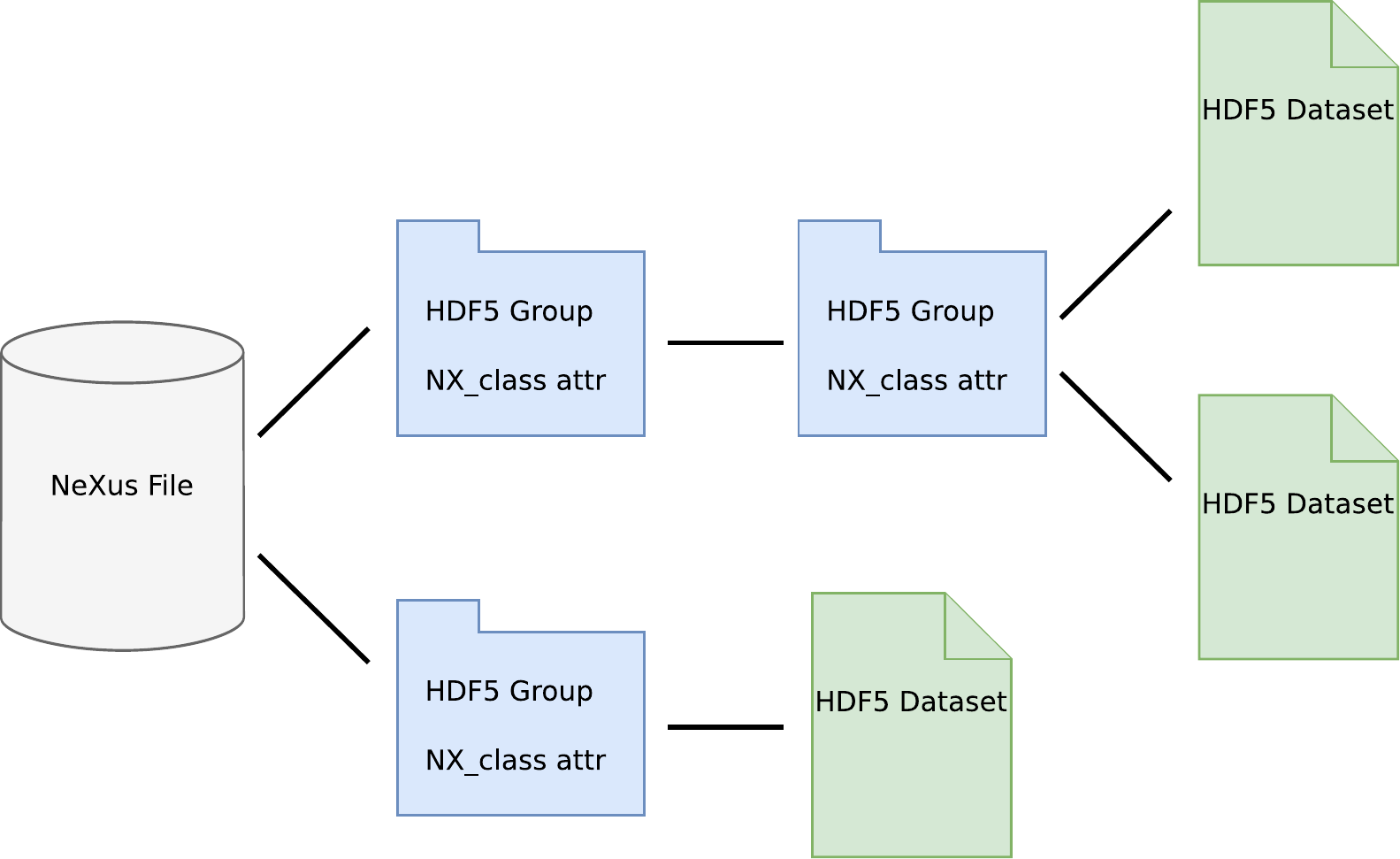}
  \caption{Diagram showing the hierarchy of a NeXus file, which comprises HDF5
    datasets and groups that conform to NeXus class definitions.}
  \label{fig:nexus_arch}
\end{figure}

The NeXus format defines a collection of classes, each of which has a set of
required or optional child datasets, classes and attributes. HDF5 groups in a
NeXus file declare their conformance to a NeXus class with an attribute called
\texttt{NX\_class}, which has a value of the name of the class.
Figure~\ref{fig:nexus_arch} depicts this for a simple hierarchy. Classes exist
to support a wide range of different data \cite{nexus:classes}, including
images, histograms and time series data. They can also describe details about
how the experiment was carried out, such as who carried out the experiment
and when, information about the sample being studied, what components comprise
the instrument, the instrument's geometry, and any other data required for the
analysis of an experiment at a neutron source or synchrotron facility.

As mentioned in Section \ref{sec:event}, key data types for the ESS are neutron
detection events and time series data from choppers or measurements of the
sample environment. These types of data are well supported in NeXus by the
\texttt{NXevent\_data} and \texttt{NXlog} classes, respectively. Each of these
classes contain datasets of values and corresponding timestamps. These
\emph{time-value} datasets are supplemented by \emph{cue} datasets which record
the index in the value and timestamp datasets at particular times. If only a
subset of the time range recorded in the value-timestamp datasets is required to be
read from file, this can be achieved by reading the smaller cue datasets and
using the indices into the large datasets. This is therefore possible without
having to read the entire time-value datasets, which may be slow or impossible
in the case that they exceed available memory. This system is also particularly
well suited to streamed data, as neutron detection events are batched in
messages and a cue can be recorded for each message.

The file writer writes NeXus files upon requests received
in the form of JSON commands via a Kafka topic. The request contains the
structure of the NeXus file which is to be written. This description of the
NeXus structure can contain the HDF5 group hierarchy, attributes, datasets
which are already known at the time when the command is sent, and
placeholders for datasets which
are to be written as the corresponding data messages appear on configured Kafka
topics. Because the file writer has access to historic data through the Kafka
cluster,
the commands can contain timestamps for the time range of the data to be written,
which can be in the past or in the future. If start and stop times are not
defined, the writing of a file starts as soon as the command is received by the
NeXus File Writer and it stops on a corresponding stop command.

The data streams, which are specified in form of the NeXus structure in the JSON
command, are handled by specified HDF5 writer modules. They process the messages
that arrive on the configured Kafka topic and are responsible for any additional
processing and the actual writing of that data to the HDF5 file. This modular
design allows easily extending the file writer to cater for new FlatBuffers
schemas, for more specialised output formats or optimised ways of writing data
to file.

\section{Tests and results}
\label{sec:results}


To verify and demonstrate the viability of the proposed approach to data
aggregation and streaming, a series of tests have been performed, covering the
integration of the components and their performance. Some of the tests are
automated and run as new versions of the applications are created, in a local
computing infrastructure with virtual machines at the DMSC; others are run on
demand on dedicated hardware. Furthermore, the system has been deployed to
operating neutron facilities for evaluation. In the following sections, the
different types of tests and their results are presented, along with a
discussion of the
deployments of the system to operating facilities.

\subsection{Integration tests}

\begin{figure}[tb]
  \centering
  \includegraphics[width=0.9\textwidth]{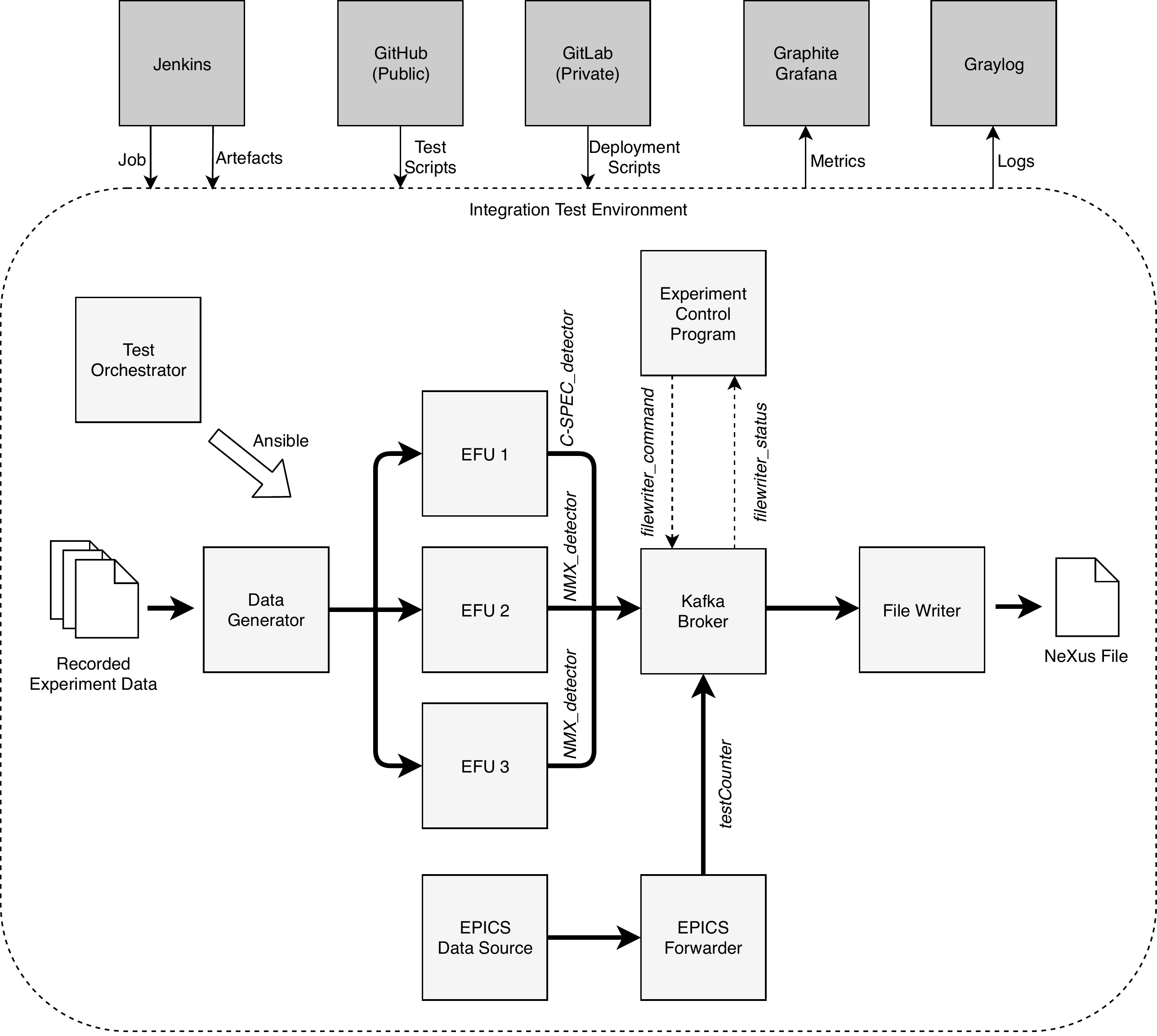}
  \caption{Integration test setup. The Jenkins build server triggers the test,
    which runs in virtual machines in the DMSC computing infrastructure (depicted
    in light grey). The test orchestrator node uses Ansible to connect to the
    other virtual machines and run the required services and commands. Data flow
    is represented by bold lines, status and commands by dashed lines, and topic
    names are written in italics.}
  \label{fig:integration_test}

\end{figure}

An automated integration testing regime including the data aggregation and streaming
software is in place in the software build and test infrastructure at DMSC,
consisting of a Jenkins continuous integration server using the
configuration management and orchestration tool Ansible. A
test is triggered every time changes are made to the code
bases of the Event Formation Unit, EPICS to Kafka Forwarder and NeXus File
Writer.
Raw data from detector
prototype measurements are streamed to the EFUs, which process the
signals into
events and forward them to the Kafka cluster. A simulation EPICS server is used to
generate process variable data for the EPICS forwarder, and the file writer is
started and configured to write detector data and EPICS metadata to a NeXus
file. Figure~\ref{fig:integration_test} shows the current setup; results are
made available through the Jenkins server web interface, while metrics and logs
can be visualised with Grafana and Graylog, respectively.

The integration test verifies the correct number of events have been sent from the
EFUs. Metrics and log data are stored for the test runs and provide
benchmark information
such as Kafka data rates. A script verifies the data
written to the NeXus file.


At the moment three dedicated servers are available at an ESS integration laboratory, where
other beamline equipment such as motion control and sample environment devices
are also located for testing purposes, along with detector prototypes. These
servers are operated for manual tests and have been used for a successful
demonstration of streaming event data from EFUs into Kafka with live
visualisation. Work is in progress to integrate the Experiment Control Program
and Fast Sample Environment application with the devices in the laboratory, and
this will allow those components to be tested with the data aggregation and
streaming system.

\subsection{Performance tests}
\label{sec:perf}


The performance of the aggregation and streaming system, including the NeXus
File Writer, was investigated at PSI using four Intel Xeon E5-2697 v4
\SI{2.60}{GHz} servers, equipped with \SI{252}{GB} of RAM and a GPFS file system
connected via 4x Infiniband FDR. This system is limited by the fact that all the
nodes share the same I/O interface; furthermore, the same interface is used for
communications to achieve optimal communication speed during the tests. On this system, a filesystem write performance test using 4 threads resulted in a \SI{5.8}{GB/s} rate, while internode network communication over Infiniband tests using 4 threads showed bandwidths above \SI{3}{GB/s} for message sizes greater than \SI{0.1}{MB}.

\begin{figure}[tbp]
  \centering
  \includegraphics[scale=0.7]{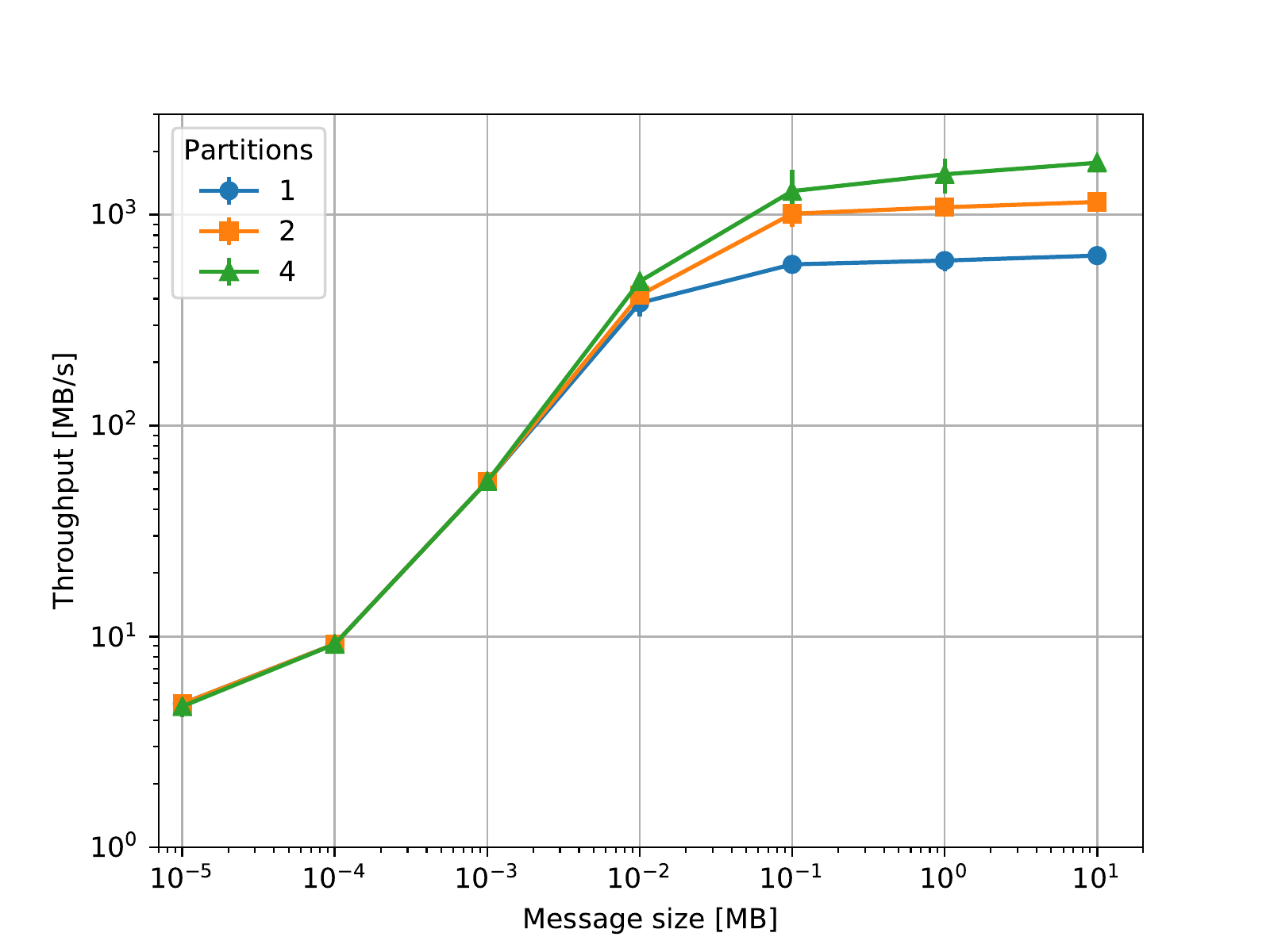}
  \caption{Kafka throughput as a function of the message size and the number of
    partitions. Two servers run the brokers, while a third server runs the event
    producer.}
  \label{fig:kafka_perf}
\end{figure}

Tests were performed using a virtual implementation of the PSI
AMOR~\cite{performance:AMOR} instrument as data source.
The output of a real measurement is
converted into a neutron event stream and sent to a Kafka cluster; the number of
events recorded in a single message can be varied in order to test the
dependence of the system on the message size. A different number of brokers and
partitions were used. Figure~\ref{fig:kafka_perf} shows the producer
performance achieved using four Kafka brokers as a function of the message size
and the number of partitions. The figure demonstrates that the overhead of
using many small messages degrades the performance; this can be acceptable for
infrequent metadata updates, but the bulk of the neutron data needs to be
sent in larger packets, as done by the EFUs, to reach the required throughput.
This system scales well with
the number of producers when using different topics and
the aggregate throughput of two producers is roughly double what is shown in
the figure.

\begin{figure}[tbp]
  \centering
  \includegraphics[scale=0.7]{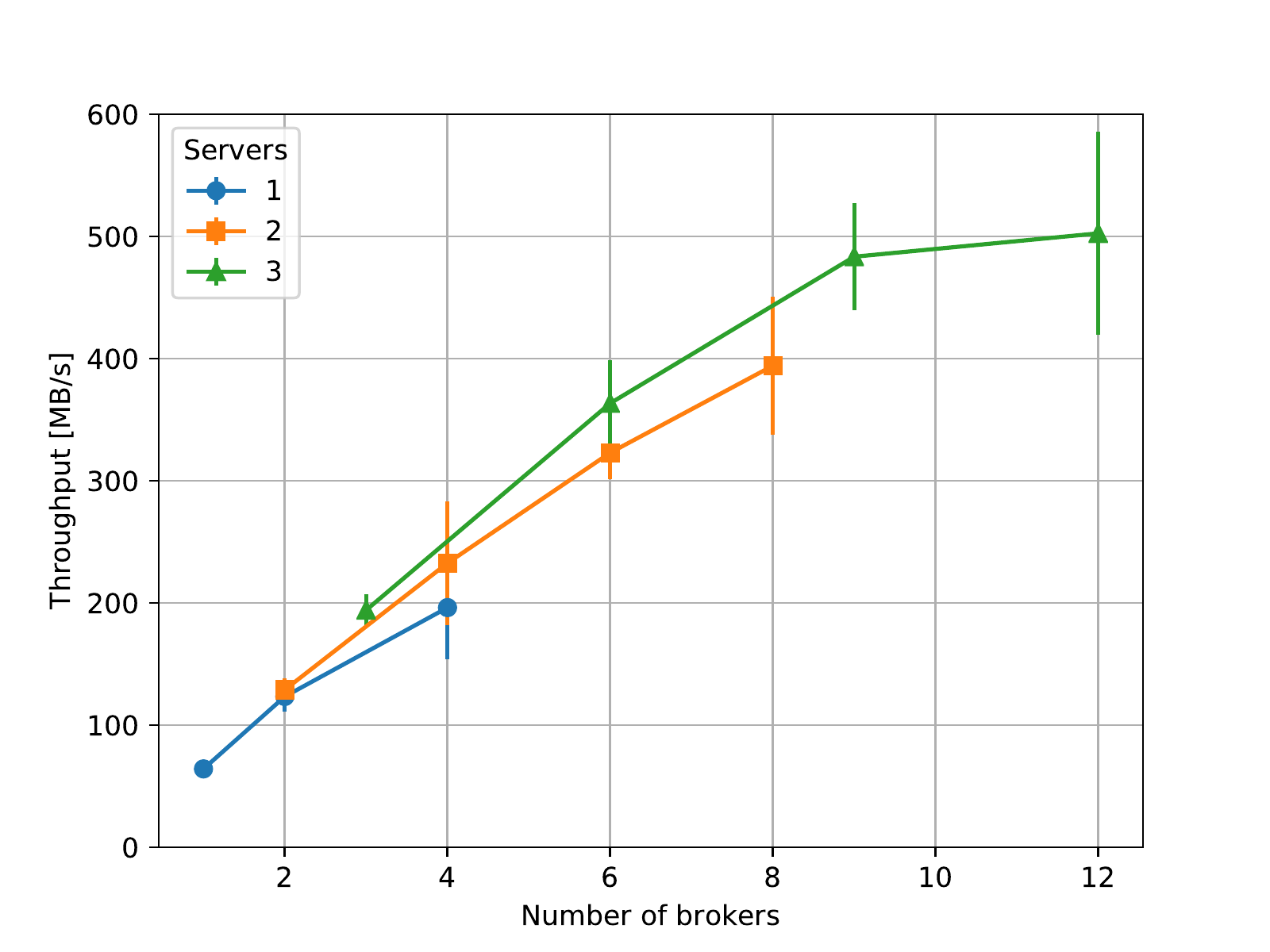}
  \caption{Scalability measurements: throughput as a function of the number of
    Kafka brokers with an event producer and a file writer, using one partition
    per broker.}
  \label{fig:kafka_scalability}
\end{figure}

Figure \ref{fig:kafka_scalability} shows the results of scalability tests with a
producer, the Kafka cluster and a file writer. For a different number of
servers, the total throughput was measured varying the number of brokers, using
one partition per broker and a \SI{1}{MB} message size.
The file writer runs on a separate server, but in the setup with three
broker machines, the producer shares the server with (at least) one Kafka broker.
This may have a negative effect on the performance on the three server case (green line).
Otherwise the data suggests a very linear scaling of the achievable throughput with
the number of deployed servers or broker processes and in turns verifies the
arguments made in \ref{sec:kafka} for choosing Kafka: that it enables an
incremental growth of the infrastructure with raising need of the
scientific instruments as ESS builds up, always keeping up with the requirements.

The performance of the HDF5 writing component is essentially limited by the
available I/O. To be able to write HDF5 files at the expected high data rates,
the file system layer has to be backed by an appropriately fast I/O layer.
Furthermore, the File Writer exposes some parameters for fine tuning
in order to achieve the best performance. On a server with Kafka messages preloaded in RAM, a single process
approach sustains about \SI{1.2}{GB/s}, while a parallelised implementation using Message Passing Interface (MPI) \cite{website:mpi} can reach
about \SI{4.8}{GB/s}; a broad discussion on the writing performances can be
found in~\cite{forwarderfilewriter:conf}, including the effects of chunking,
message size and buffering.

\subsection{Deployments at operating facilities}
\label{sec:deployments}


The data aggregation and streaming system was deployed in December 2017 to the
V20 beamline at the Helmholtz-Zentrum Berlin's BER II research reactor. That
deployment comprises a Kafka broker, the EPICS to Kafka Forwarder and the NeXus
File Writer, along with applications for obtaining system metrics. Running the
software infrastructure there allowed debugging the software and its
configuration. An ESS mini-chopper was installed in the beamline and the system
was able to obtain its data via EPICS and write them to file.


Another current deployment of the system is running at the STFC's ISIS Neutron
and Muon Source. Three instruments are using the EPICS to Kafka Forwarder
and a Kafka cluster. The system architecture differs from the
ESS architecture in that the detector data do not come from EFUs, but from data
acquisition electronics through the Instrument Control Program (ICP), together
with metadata; moreover, the ICP is responsible for writing NeXus files. The EPICS forwarder is used to send beamline metadata to Kafka,
while the Mantid Live Listener
consumes live data from the cluster. Grafana is also being used for metrics visualisation. In two of the instruments, the system is
being run in parallel with the current facility system, but in one of them,
which is a new instrument, the Kafka-enabled ICP is the main control program.
The ISIS Kafka cluster runs on three servers with Xeon R5-2620 \SI{2.10}{GHz} 8-core processors, \SI{32}{GB} of RAM and three \SI{1.2}{TB} disks in RAID~0 (for
each server).

\section{Conclusion and future work}
\label{sec:conclusion}


ESS will use a system architecture based on data aggregation and streaming for
acquisition of neutron data in event mode and EPICS metadata, all of which will
be timestamped at their source using the ESS timing system. An Apache Kafka
cluster connects the applications producing and consuming these data, including
Event Formation Units, Fast Sample Environment devices, EPICS Input/Output
Controllers using the EPICS to Kafka Forwarder, Mantid and the NeXus File
Writer. The solution can be tuned for balancing the throughput, latency and
storage concerns arising from requirements, and also provides scalability and
availability through partitioning and replication.

The Google FlatBuffers library is used for data serialisation and, with the
schema definitions in a common code repository under source control, decouples
the producer and consumer applications. This decoupling means parts of the
system can be substituted by alternative components that use the appropriate
schemas, making the system more generic and also usable at other neutron
facilities.

Integration and performance tests have been run and demonstrate the viability of
the solution for ESS, showing the desired functionality and scalability
behaviour. Deployments to operating neutron facilities, like the
Helmholtz-Zentrum Berlin's BER II research reactor and the ISIS Neutron and Muon
Source, corroborate this.

In the future the integration of the forwarder and file writer
applications with the Experiment
Control Program will be improved and the whole system tested for robustness
in a number of failure scenarios.
End-to-end tests will be performed at the integration
laboratory at ESS, using real beamline equipment and detector readout
systems, as well as at the V20 beamline at HZB.

\acknowledgments

This work is partially funded by the European Union Framework Programme for
Research and Innovation Horizon 2020, under grant agreement 676548.


\end{document}